\documentclass[twocolumn]{aastex63}

\usepackage{graphicx,epstopdf,amsmath}
\usepackage{import}

\usepackage{color}

\newcommand{\natph}{NatPh}
\newcommand{\phpl}{PhPl}

\begin{document}

\title{Is flare-ribbon fine structure related to tearing in the flare current sheet?}


\author{P.~F.~Wyper} 
\affil{Department of Mathematical Sciences, Durham University, Durham, DH1 3LE, UK}
\email{peter.f.wyper@durham.ac.uk}

\author{D.~I.~Pontin} 
\affil{School of Mathematical and Physical Sciences, University of Newcastle, University Drive, Callaghan, NSW 2308, Australia }
\affil{School of Science and Engineering, University of Dundee, Dundee, DD1 4HN, UK}
\email{david.pontin@newcastle.edu.au}

\begin{abstract}
Observations of solar flare ribbons show significant fine structure in the form of breaking wave-like perturbations and spirals. The origin of this structure is not well understood, but one possibility is that it is related to the tearing instability in the flare current sheet. Here we study this connection by constructing an analytical three-dimensional magnetic field representative of an erupting flux rope with a flare current sheet below it. {We introduce small-scale flux ropes representative of those formed during a tearing instability in the current layer, and use the squashing factor on the solar surface to identify} the shape of the presumed flare ribbons and fine structure. Our analysis suggests there is a direct link between flare-ribbon fine structure and flare current sheet tearing, with the majority of the ribbon fine structure related to oblique tearing modes. Depending upon the size, location and twist of the small-scale flux ropes, breaking wave-like and spiral features within the hooks and straight sections of the flare ribbon can be formed that are qualitatively similar to observations. We also show that the {handedness} of the spirals/waves must be {the same as the handedness of the} hooks of the main ribbon. We conclude that tearing in the flare current layer is a likely explanation for spirals and wave-like features in flare ribbons.
\end{abstract}



\section{Introduction}\label{sec:intro}
Flare ribbons below erupting coronal structures are understood to be the chromospheric footprint of flare reconnection in the corona \citep[e.g.][]{Shibata2011,benz2017}, with the surface flux swept out by flare ribbons being directly related to the magnetic flux reconnected through the flare current layer in that time  \citep[e.g.][]{forbes2000}. Since direct imaging of flare reconnection is highly challenging, flare ribbons are therefore often used as a diagnostic for the flare reconnection process \citep[e.g.][]{Wang2003,Kazachenko2017}.

Early 2D conceptual models of eruptive flares envisaged reconnection occurring at an X-point beneath an erupting flux rope (O-point in 2D), forming two parallel ribbons either side of the PIL; the CSHKP model \citep{Carmichael1964, Sturrock1966, Hirayama1974, Kopp1976}. Later conceptual models then incorporated the finite length of the erupting structure and the strong guide field present within the filament channel \citep[e.g.][]{Moore2001,Priest2002}. More recently the 3D evolution of flare reconnection and its association with flare ribbons has been put on a firmer theoretical footing through numerous numerical experiments and advances in our understanding of 3D topology and reconnection in general \citep[e.g.][]{Longcope2007,Isenberg2007,aulanier2012,aulanier2013,janvier2013,Wyper2021}. 

 \begin{figure}
   \centering
      \includegraphics[width=1\columnwidth]{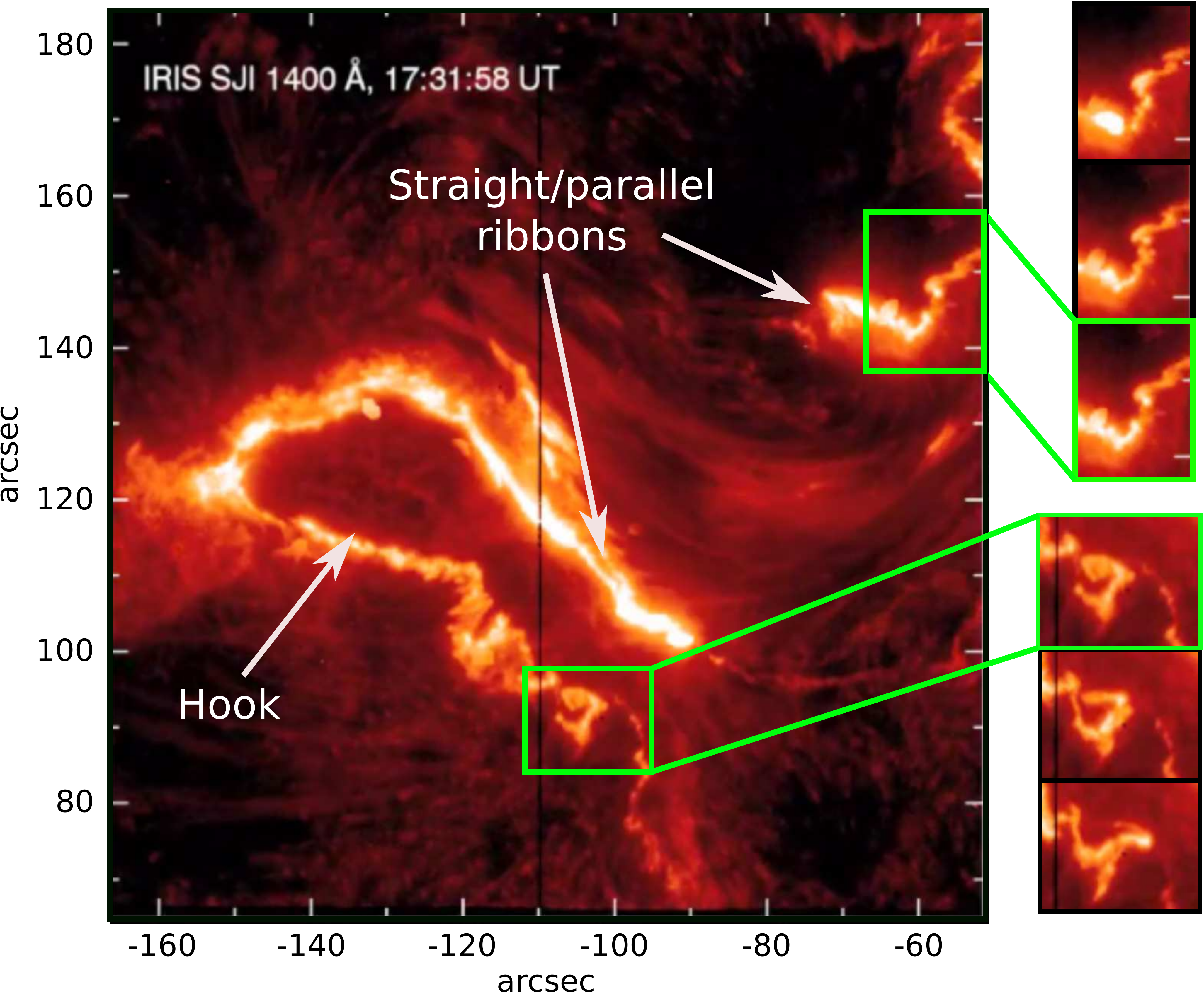}
   \caption{{\it{Interface Region Imaging
Spectrograph}} (IRIS) observation of the 2014 September X-class flare that exhibited spiral and wave-like motions of the flare ribbons.}
\label{fig:iris}
    \end{figure}

In 3D the X-point within the flare current layer is replaced with a Hyperbolic Flux Tube (HFT) beneath a flux rope anchored to the surface at both ends. Depending upon the eruption trigger scenario the HFT/flux rope pair can be pre-existing (e.g. Torus Instability; \citet{Kliem2006}) or form during the eruption (e.g. Magnetic Breakout; \citet{Antiochos1999}) and in general is likely to exist in a state of flux between the two \citep{Patsourakos2020}. However, most 3D eruption scenarios agree on this general topology once flare reconnection is initiated. 

As the flux rope is anchored at both ends the change in the field line mapping is rapid, but not now discontinuous as one considers the connectivity of foot points moving from the flare loops to the overlying arcade, or the overlying arcade to the flux rope \citep[e.g.][]{aulanier2012}. This quasi-boundary between these structures is formed by quasi-separatrix layers \citep[QSLs;][]{Titov2007}, with the HFT formed by the intersection of two sheet-like QSLs that wrap around the flux rope \citep{Titov1999}. The footprints of these two QSLs form two ribbon-like straight/parallel sections beneath the flux rope, each with a J-shaped hook at their end, {outlining the intersection of the flux rope with the surface}. The HFT sits at the centre of the flare current layer and so heat flux and non-thermal particles from the flare reconnection follow the QSL field lines down to the solar surface. The footprints of QSLs calculated on the solar surface have therefore become a powerful tool for understanding the morphology of flare ribbons in eruptive flares \citep[e.g.][]{Savcheva2015,Janvier2016,Zhao2016}. 

There is mounting evidence from both theory and observations that reconnection within the flare current layer is fragmented and bursty in nature. 
Observations of plasmoids in post CME rays, hard X-ray bursts, and intermittent downflows \citep[e.g][]{McKenzie1999,Riley2007,Kliem2000,Asai2004,Cheng2018} all suggest that flare reconnection is a fundamentally fragmented and bursty process. Linear theory and numerical {{reconnection experiments}} show that for the high Lundquist numbers characteristic of the solar corona, high aspect ratio current layers form which then rapidly become unstable to tearing \citep[e.g.][]{Loureiro2007,Bhattacharjee2009,Edmondson2010,Pucci2014,Wyper2014a}. In 2D this produces multiple magnetic islands with properties that follow power laws \citep[e.g.][]{Huang2010}. {{Indeed, motivated by early reconnection experiments \citet{Shibata2001} introduced the idea of ``fractal reconnnection'' within the flare current layer whereby repeated tearing and current sheet thinning leads to a fractal-like distribution of plasmoids and current sheets. Plasmoid formation, sometimes in this fractal-like manner, is a common feature of highly resolved flare reconnection in 2.5D CME simulations \citep[e.g.][]{Barta2011,Karpen2012,Lynch2013,Lynch2016b,Guidoni2016,Hosteaux2018}}}. 

{{However, when extended to 3D the dynamics are considerably more complex. Without a guide field, short, dynamic plasmoids with highly twisted field lines form and evolve in a fully 3D manner \citep{Edmondson2010,Nishida2013}. With the inclusion of a guide field linear theory and numerical experiments show that oblique modes can form on multiple flux surfaces within the current layer \citep[e.g.][]{Daughton2011,Baalrud2012,Wyper2014b,Huang2016,Edmondson2017,Stanier2019}. Such oblique modes form flux ropes at an angle to the guide field which in the non-linear phase overlap and interact, sometimes leading to a turbulent cascade \citep[e.g.][]{Huang2016}. Fully 3D CME simulations with a realistically evolving guide field are only now beginning to reach Lundquist numbers where such plasmoids are resolvable.}}

Flare ribbons provide a further potential piece of indirect evidence of the fragmented/turbulent nature of flare reconnection. Flare ribbons often have multiple kernels and generally exhibit a complex structure and evolution, especially when viewed in close detail \citep[e.g.][]{Asai2002,krucker2003,Brannon2015,Jing2016,Li2015,Li2018}. For instance, \citet{Brannon2015} analysed bright knots and wave-like perturbations in a section of flare ribbon and \citet{Parker2017} later tried to explain these findings based on a quasi-two-dimensional tearing analysis involving velocity shear flows. Aside from wave-like perturbations, spirals are also occasionally observed in flare ribbons. An example is shown in Fig. \ref{fig:iris} where an evolving spiral in the hook and wave-like evolution of a straight section of the ribbon are highlighted. \citet{Dudik2016} studied this event in detail \citep[see also e.g.][]{Cheng2015,Li2015,Zhao2016} and noted that the hook especially continually evolved (``squirmed'') with similar spiral structures. Although in the past spiral structures have been attributed to the Kelvin-Helmholtz instability \citep{Ofman2011}, in our view these are the most compelling signature of flux rope formation within the current layer. 

In our previous work on the fragmentation of current sheets at 3D null points, we have shown that the small-scale flux ropes/plasmoids that form due to tearing wrap up the separatrix surface which then maps to spirals on the surface (or boundary of the domain) \citep{Wyper2014b,Pontin2015}. {{The presence of such plasmoids has been confirmed by high resolution simulations and observations of null point reconnection in coronal jets }} \citep{Moreno-Insertis2013,Wyper2016b,Kumar2018,Kumar2019}. Moreover, studies of non-thermal particle acceleration indicate that the photospheric particle impact patterns are also sensitive to the formation of the plasmoids, being guided along the flux ropes that form the plasmoids \citep{Pallister2019,Pallister2021}. All of the above suggest that {{similar tearing-induced structure within a flare}} current layer is a strong candidate for explaining flare ribbon fine structure in eruptive two-ribbon flares. 

In this work we explore this idea. {{Rather than simulating the self consistent formation of a 3D flare current layer and its associated fragmentation, which is a formidable task, we attack the problem analytically and}} consider a simple model magnetic field that contains all the expected topological ingredients of a flux rope eruption: a 3D flux rope above a current sheet formed at an HFT. The simplicity of the model affords us control of where and in what way we simulate tearing, which we do by introducing local regions of twist to form small-scale flux ropes. We find that flare ribbon spirals/wave-like motions are an expected feature of tearing in the current layer, but only when the flux ropes are oblique. We conclude that flare ribbon structure is indeed likely a result of tearing and flux ropes formed in the flare reconnection region.

\begin{figure*}
\centering
\includegraphics[width=2\columnwidth]{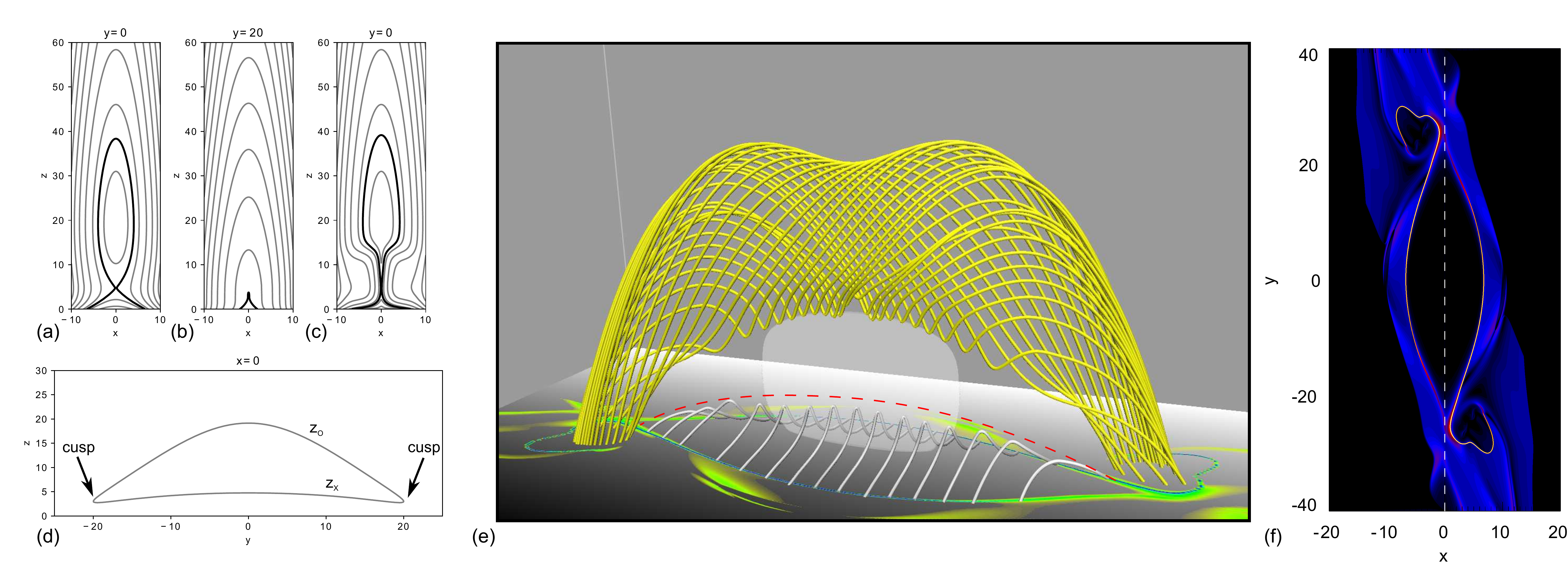}
\caption{Contours of the vector potential without the current sheet (a) and (b), and with it (c). Thick black lines outline the separatrices in these 2D projections, which closely match the QSLs at $y=0$. (d) shows $B_x=0$ in the $x=0$ plane in both fields. (e) field lines showing the core of the erupting flux rope (yellow), flare loops (silver) and the flare current layer (semi-transparent isosurface). $\log(Q)$ is shown on the surface in yellow/green. The dashed red line shows the approximate position of the HFT. (f)  $\log(Q)$ on the surface ($z=0$) scaled to the maximum value. The dashed line shows the PIL.}
\label{fig:vect}
\end{figure*}

\section{Methods}
\subsection{Background Field}
Our background field is given by
\begin{equation}
\mathbf{B}_{b} = \mathbf{B}_{fr}+\mathbf{B}_{cs}
\end{equation}
where $\mathbf{B}_{fr}$ produces a large-scale flux rope and HFT, and  $\mathbf{B}_{cs}$ a current layer. The flux rope field takes the form
\begin{align}
\mathbf{B}_{fr} &= \boldsymbol{\nabla}\times (A_0  \hat{\mathbf{y}}) + b_0 \hat{\mathbf{y}},
\end{align}
where 
\begin{align}
A_0 &= \frac{x^2}{2} + z \nonumber\\
&+\frac{16(z-z_0)}{\left(1+y^2/L_y^2\right)\left(1+(z-z_0)^2/L_z(y)^2\right)}\nonumber\\
L_{z}(y) &= \beta [1-(y/L_y)^2]+\gamma, \nonumber\\
z_{0} &= z_{min}-\sqrt{3}\gamma.
\label{eqn:vect}
\end{align}
Throughout the majority of this work we set $L_y = 20$, $\gamma = 3$, $z_{min} = 5$ and $b_0 = 1.7$. This is a generalisation of the field first used by \citet{Hesse2005} and then further explored by \citet{Titov2009}, see Appendix \ref{ap:A} for further details. At $y=0$, the contours of $A_0$ form an O-point above an X-point, see Fig. \ref{fig:vect}(a). (Note that these contours do not strictly show field lines but they do give an idea of the local field structure projected onto this plane.). As $|y|$ increases, the X and O-points approach each other; eventually merging at a cusp point when $|y|=L_y$ (Fig. \ref{fig:vect}(b)). The $z$ positions of the $X$ and $O$ points (corresponding to where $B_x(0,y,z)=0$) are shown in Fig. \ref{fig:vect}(d) and are given by 
\begin{align}
z_{O,X} &= z_{0} \nonumber\\
&+L_{z}\left\{ \pm\left[ \frac{32}{\zeta(y)}\left(\frac{2}{\zeta(y)} -1\right) \right]^{1/2} + \frac{8}{\zeta(y)}-1 \right\}^{1/2} ,
\end{align}
where $\zeta(y) = 1+y^2/L_{y}^2$ (see Appendix \ref{ap:A}, for the general expression). Near $y=0$, $z_O$ and $z_X$ closely follow the centre of the flux rope and the HFT, respectively. 

To form a strong current layer at the HFT consistent with the simulations and observations discussed in the previous section, $\mathbf{B}_{cs}$ takes the form 
\begin{align}
\mathbf{B}_{cs} =  
\begin{cases} 
\nabla \times \left[ c_{1} f(x) g(y) h(y,z) \hat{\mathbf{y}}\right] & (y/L_y)^2 \le 1 \\
0 & \text{otherwise}
\end{cases}
\end{align}
where
\begin{align}
f(x) &= \ln\left[ \cosh\left( x/l_{x}\right)\right] e^{-\left(x/k_{x}\right)^2},\nonumber\\
g(y) &= \frac{[1-(y/L_y)^2]^2}{[1+(y/L_y)^2]^2}, \nonumber\\
h(y,z) &= \left[ \tanh\left(\frac{z-z_{X}(y)+z_{c}}{l_{z}}\right) - \tanh\left(\frac{z-z_{X}(y)-z_{c}}{l_{z}}\right) \right] \nonumber\\
&\times  \tanh\left( \frac{z}{k_{z}}\right).
\label{eqn:A1}
\end{align}
{{A full description of each function is given in Appendix \ref{ap:B}.}} Here we choose $c_1 = 0.6, k_x = 5, k_z = 0.2, l_x = 0.1, l_z = 2.0$ and $z_c = 6.0$. Since $\mathbf{B}_{cs}(x = 0) = \mathbf{B}_{cs}(z = 0) = 0$, the addition of this second field does not affect the positions of $z_{X}$ and $z_{O}$ or the normal magnetic field on the solar surface. Figure \ref{fig:vect}(c) shows contours of the combined flux functions at $y=0$, while Fig.  \ref{fig:vect}(e) shows a 3D volume rendering of the current layer and field lines within the flux rope, respectively. The current layer is fully 3D and stretched beneath the flux rope. The squashing factor on the surface is shown in Fig. \ref{fig:vect}(e) and (f), showing that this field has the typical J-shaped hooks and parallel strips of high-Q associated with the flare ribbons of many eruptive flares.

\subsection{Small-scale flux rope field}
To this background field we add small-scale twists which model the local conversion of magnetic shear to twist that occurs as plasmoids form within the flare current layer. Their magnetic field takes the form
\begin{equation}
\mathbf{B}_{I} = \boldsymbol{\nabla}\times( A_{I}\hat{\mathbf{y}}),
\end{equation}
where
\begin{align}
A_{I} = &\left\{c_{2}-c_{1}\ln\left[\cosh\left(x/l_{x}\right)\right]\right\} \nonumber \\
\times & \,e^{-(x-x_{I})^{2}/r_{x}^{2} -(y-y_{I})^{2}/r_{y}^{2} -(z-z_{I})^{2}/r_{z}^{2} },
\end{align}
and $c_1$ and $l_x$ are as in $\mathbf{B}_{cs}$. The $\log$ term creates a cavity within the current layer where the magnetic shear is removed, while the first term places a rotation into this cavity. {The magnetic field structure induced is consistent with those formed during the 3D tearing simulations described in Section \ref{sec:intro}.} For the oblique modes we considered small-scale flux ropes which were rotatable about $(x_I, y_I, z_I)$ in the $yz$-plane. For simplicity in this case we set $r_y = r_z = r$, so that the perturbation field takes the form: 
\begin{equation}
\mathbf{B}_{I} = \boldsymbol{\nabla}\times[\cos(\theta) A_{I} \hat{\mathbf{y}} + \sin(\theta) A_{I} \hat{\mathbf{z}}] 
\end{equation}
where $\theta$ is the angle the axis of rotation makes with the horizontal direction. 

The final magnetic field is given by a combination of the background, current sheet and island fields:
\begin{equation}
\mathbf{B} = \mathbf{B}_{fr} + \mathbf{B}_{cs} + \sum_{m=1}^{n} \mathbf{B}_{I,m},
\end{equation}
where $n$ is the number of islands.

     \begin{figure*}
   \centering
   \includegraphics[width=2\columnwidth]{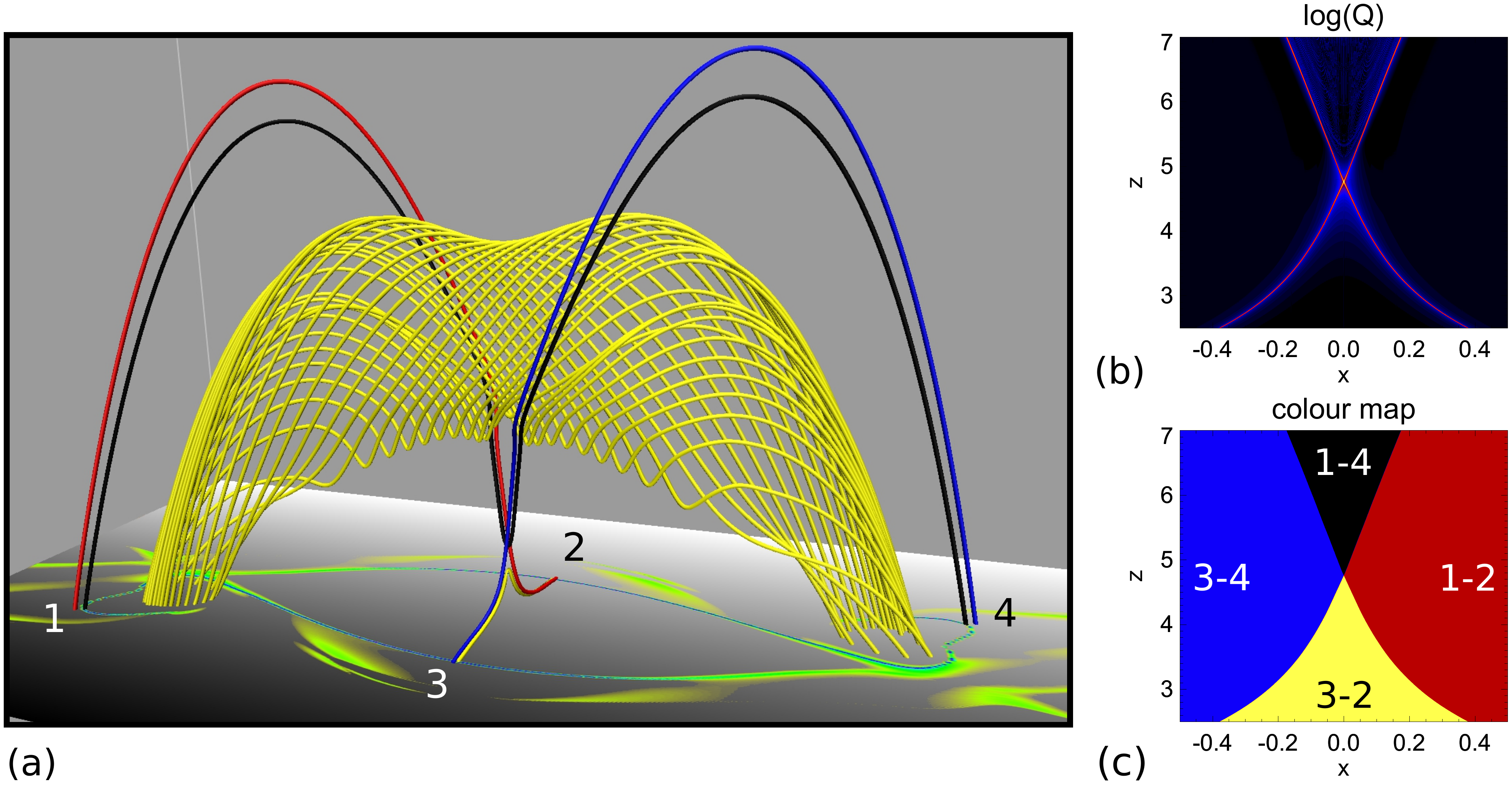}
   \caption{(a) {{Representative}} field lines showing the four main {{connectivity types present in the field ${\bf B}={\bf B}_{fr}+{\bf B}_{cs}$}}. Red and blue: arcade field lines {{on the flanks, in what would be the reconnection inflow region in the standard two-ribbon flare picture}}. Black: field lines in the sheath around the main flux rope. Lower yellow: flare loop field lines. (b) log(Q) in the mid-plane ($y=0$) at the HFT where the four field line regions meet. (c) colour map showing the different connectivities. The colours are  matched to the field lines in (a).}
              \label{fig:index}
    \end{figure*}

 \begin{figure*}
   \centering
   \includegraphics[width=2\columnwidth]{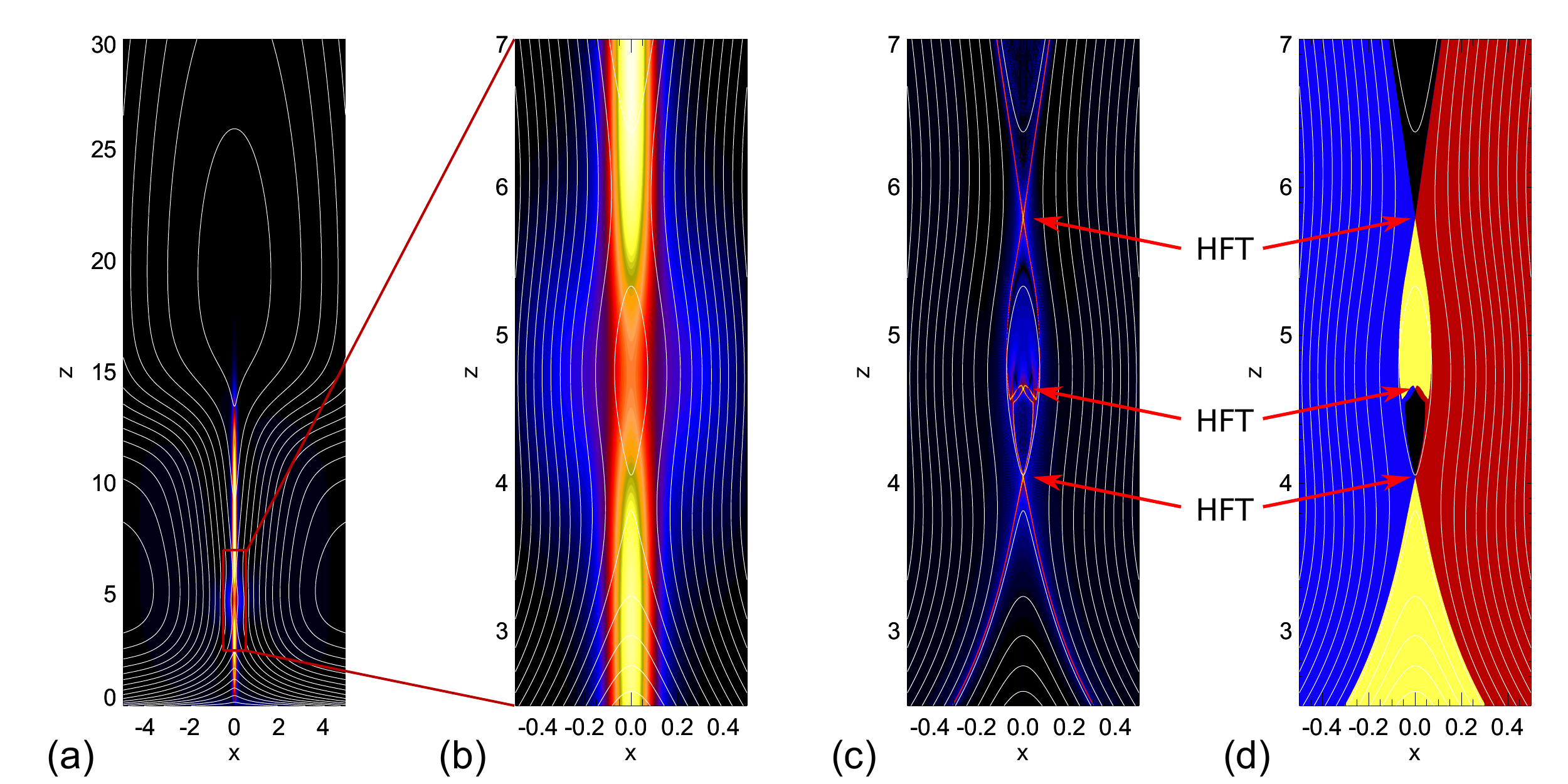}
   \caption{Island centred on the HFT with $c_{1} = 1.0$. (a) wide view of current in mid-plane ($y=0$). Contour lines show the $y$-component of the vector potential in the plane. The box shows where the other panels are taken from. (b) current. (c) squashing factor $\log(Q)$ outlining the boundaries between the different flux systems. (d) colour map showing the connectivity of each region. }
              \label{fig:island}
    \end{figure*}

 \begin{figure}
   \centering
   \includegraphics[width=\columnwidth]{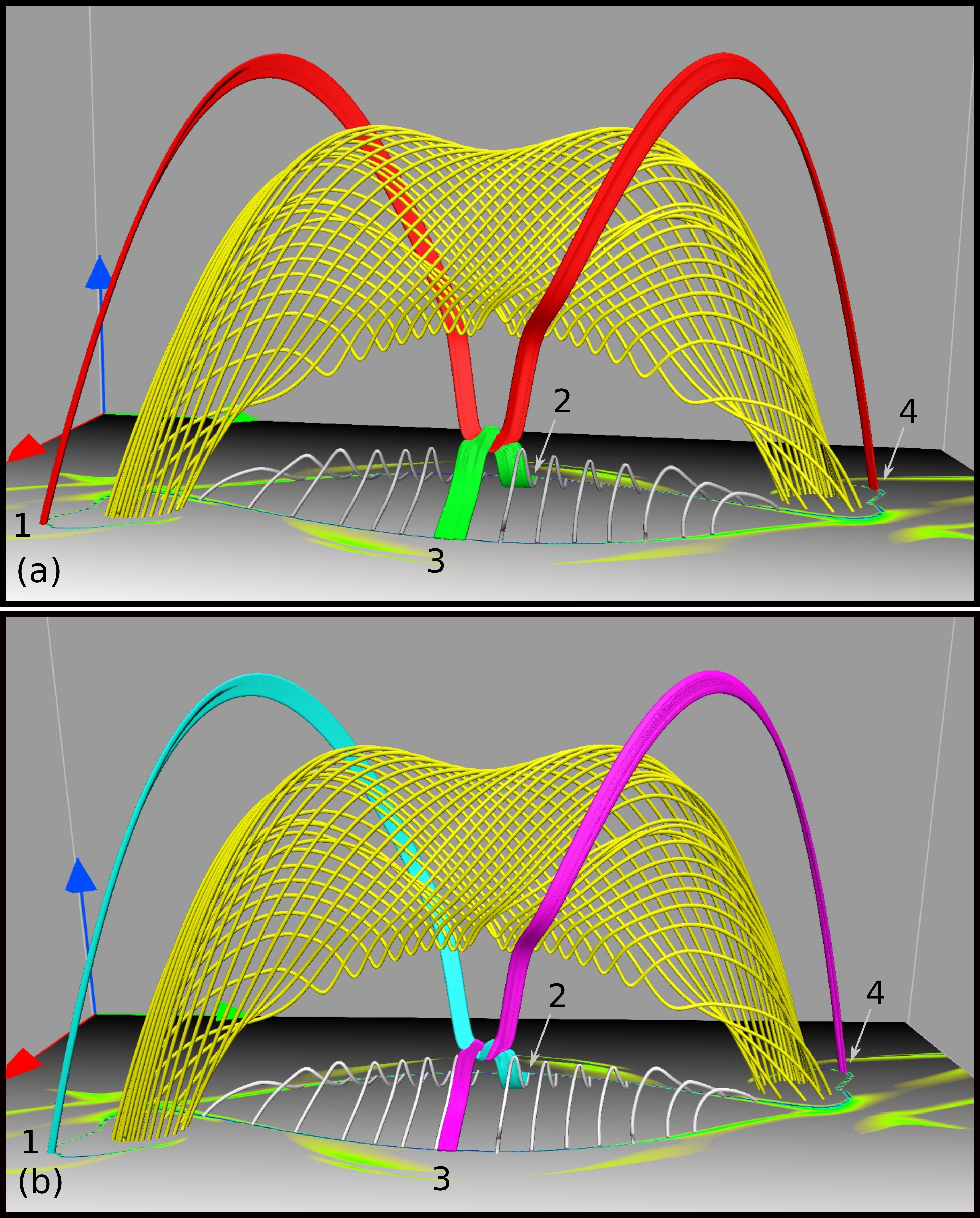}
   \caption{(a) the first flux tube pair formed by the first HFT bifurcation (red and green), $c_2 = 1.0$. (b) the second pair (cyan and magenta), $c_{2} = 1.25$. Note that the first pair are also present, but not shown for clarity.}
              \label{fig:vap_hft}
    \end{figure}

     \begin{figure*}
   \centering
   \includegraphics[width=2\columnwidth]{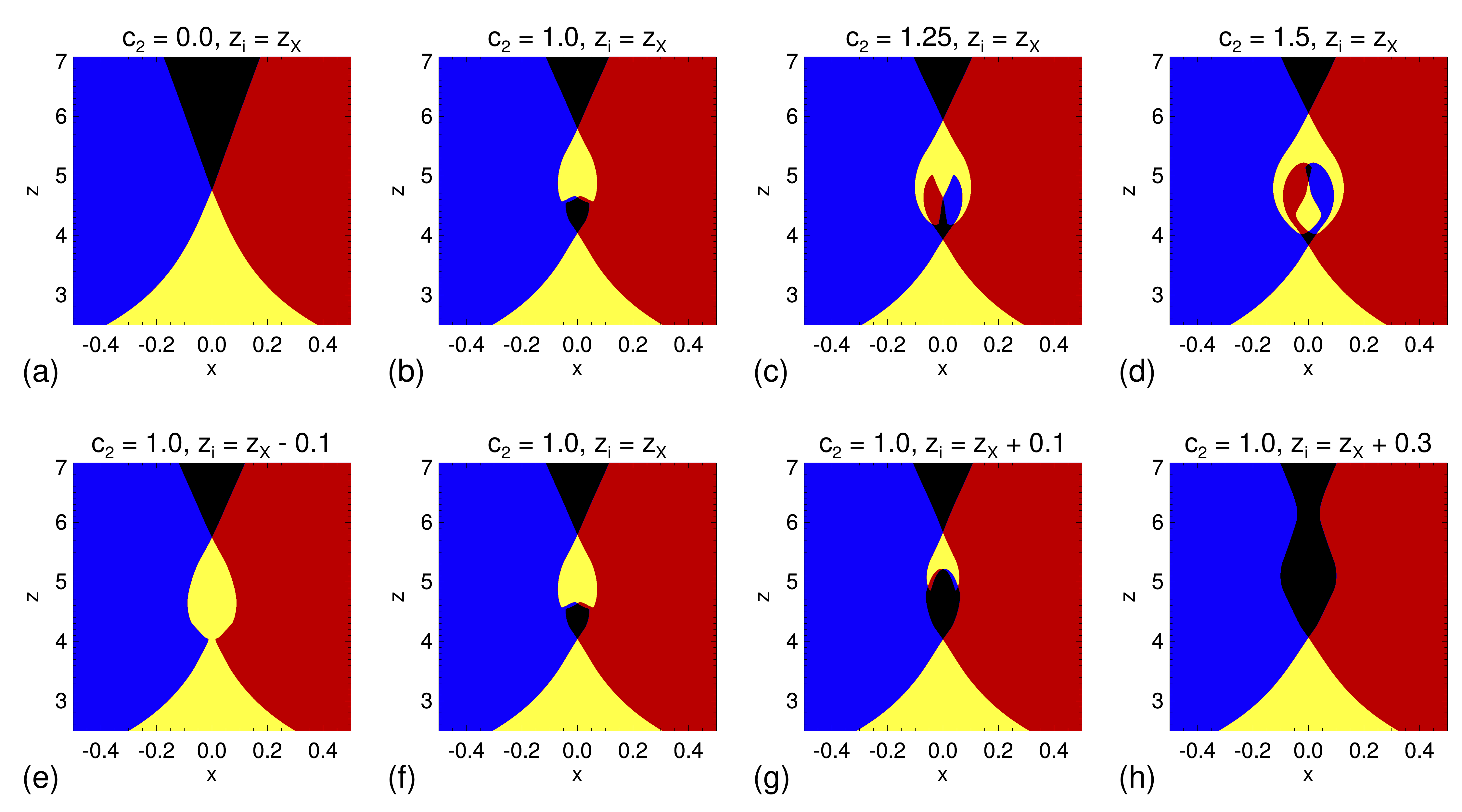}
   \caption{Colour maps showing island formation via increasing the twist of the flux rope (a)-(d), and island ejection via varying the height of the island (e)-(h).}
              \label{fig:index_comp}
    \end{figure*}

 \begin{figure}
   \centering
   \includegraphics[width=\columnwidth]{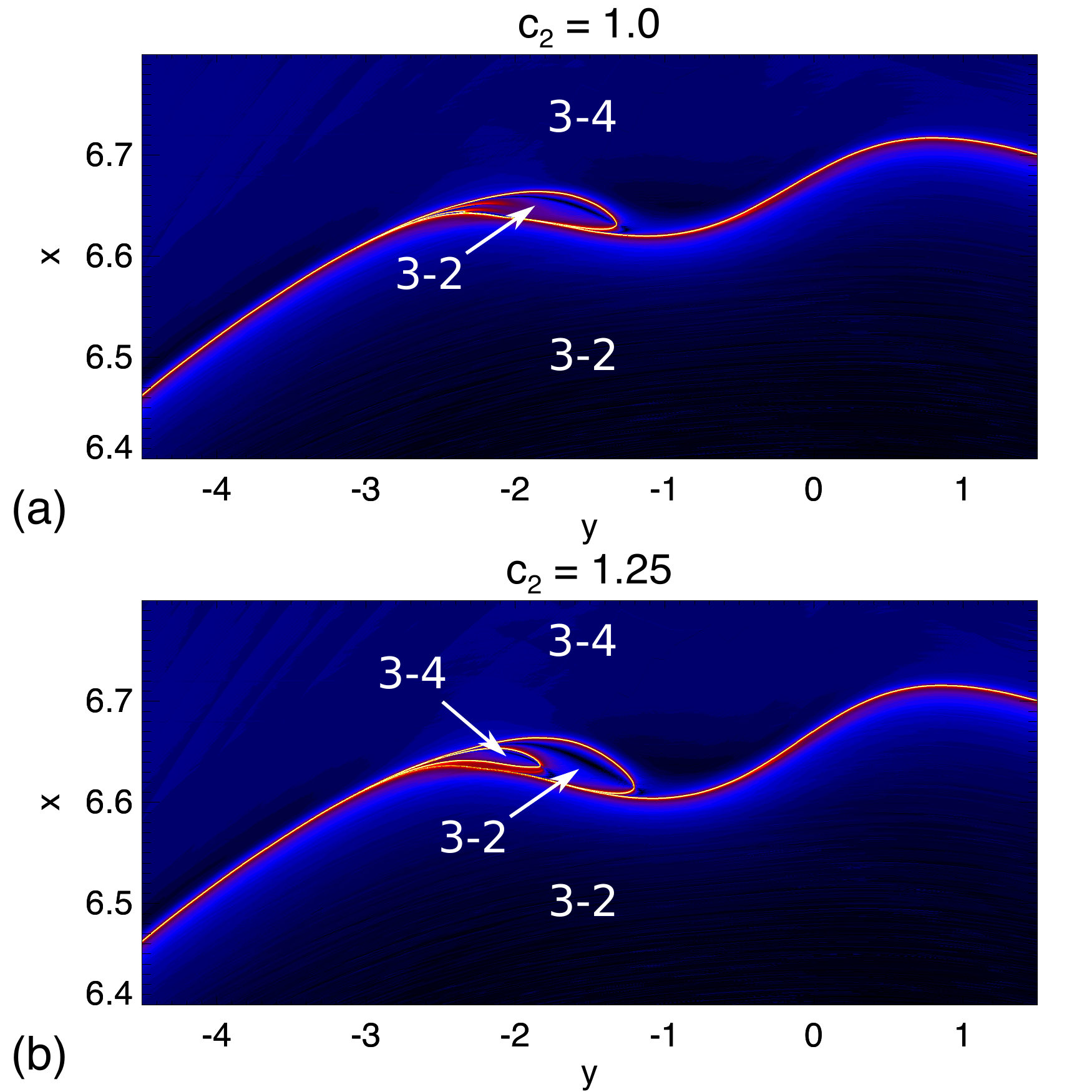}
   \caption{$\log(Q)$ showing the sub-structure formed in the straight section of the flare ribbon. (a) after the first bifurcation. (b) after the second. Foot points of arcade field lines are above the ribbon, whereas flare loop foot points are below.}
              \label{fig:hft_ends}
    \end{figure}

\section{Results}
As previously discussed, the HFT sits at the center of the flare current layer and divides {{domains of different field line {connectivity}. Field lines immediately next to the HFT in each of these domains are shown in Fig. \ref{fig:index} alongside a plot of $\log(Q)$ at the centre of the HFT (in the $y=0$ plane), Fig. \ref{fig:index}(b). To classify {the} field line {connectivity} we split the {$z=0$ plane (``photosphere'')} into four {regions}, as shown in Fig. \ref{fig:index}(a): {regions} $1$ and $4$ are {located} beyond the cusp points and near the hooks, {while} $2$ and $3$ are between the cusps, near the parallel sections of the ribbon.}} The black field line {{($1$-$4$)}} is in the sheath of flux around the core of the flux rope and loops underneath it. The yellow field line {{($3$-$2$)}} is in the flare loops, while the red {{($1$-$2$)}} and blue {{($3$-$4$)}} field lines form part of the overlying arcades. For what follows it is useful to introduce a colour map based on these four field line regions, shown in Fig. \ref{fig:index}(c) where blue, red, yellow and black correspond to regions of the plane threaded by flux corresponding to the left and right arcades, the flare loops and the flux rope sheath, respectively.

\subsection{2D-like flux ropes}
\subsubsection{Structure within the current layer}
First we consider the case most analogous to the start of tearing in a 2D setting; the formation of a single small flux rope at the center of the HFT with an axis parallel to the direction of current (the $y$ direction). Figure \ref{fig:island}(a) and (b) shows the current in the mid-plane ($y=0$), along with contours of the in-plane vector potential. Although these contours are not field lines, they would become field lines in the limit of zero guide field ($b_0=0$) and are often used in 2.5D studies to follow the flux evolution {{and are the most readily comparable to the CSHKP model and its extensions to include plasmoids.}} {{As previouly discussed,}} such studies have identified or inferred an often near-fractal formation of islands within the flare current layer with chains of X and O-points repeatedly forming and merging {{\citep[e.g.][]{Shibata2001,Barta2011,Lynch2016b}.}} 

Here the contours of the flux function depict a similar formation of an O-point along with two X-points. However, $\log(Q)$ (Fig. \ref{fig:island}(c) and (d)) reveals the true 3D evolution; that in fact the original HFT has bifurcated into three (note that any point where the four colours meet defines the axis of an HFT). {{This bifurcation is a fully 3D reconnection effect and occurs}} when field lines from the flare loop region become folded over those from the CME sheath region, which appears in the colour map  (Fig. \ref{fig:island}(c)) as the formation of a new black and yellow region bordered by each of the three HFTs.  The field lines within these new regions are shown in red and green in Fig. \ref{fig:vap_hft}(a) {{and show that part of the ``in flow'' arcade field which originally connected the surface regions $1$ to $2$ and $3$ to $4$ has been converted to sheath ($1$-$4$) and flare loop ($3$-$2$) field lines as the twist develops within the plasmoid/small flux rope.}} 

The more twisted the small flux rope that forms, the greater the number of bifurcations and subsequent wrapping regions of field lines. Fig. \ref{fig:vap_hft}(b) shows field lines in flux regions formed by the next bifurcation. These flux tubes wrap one half turn more before diverging from each other and connecting to the surface. {{They now connect to the same surface regions as the original arcade field did ($1$-$2$ and $3$-$4$). However, they are distinct from the nearby arcade field as they wrap around each other once within the plasmoid.}} {{Each subsequent bifurcation continues in this way and produces a new pair of HFTs}} alongside a pair of nested flux ropes with a half turn more twist {{and a similar jump back and forth of surface connectivity ($1$-$2 \to 1$-$4$ and $3$-$4 \to 3$-$2$ and back again)}}. Colour maps in Fig. \ref{fig:index_comp}(a) to (d) show the formation of these new nested flux regions as the twist in the flux rope is increased; recall the HFTs are at the locations where the four colours meet. 

{{The formation of these new flux tubes shows the sensitive coupling between local effects within the current layer and the global connectivity of the system. It also highlights the difficulty in interpreting a fully 3D magnetic field evolution using a 2D slice. Locally, the formation of the plasmoid is a simple local increase in twist. However, the global consequences of this are large jumps in the distant foot points that the plasmoid field lines map to as the twist of the plasmoid varies.}}

{{While our model in static, it is well established that bifurcations such as those discussed above also occur in dynamic evolutions. HFT bifurcations occurred in the MHD simulations performed by \citet{Wilmot-Smith2007}. There, as here, the field within the current layer had an O-point structure in a cross-sectional slice, while connectivity plots revealed this region contained multiple dynamically formed HFTs. Furthermore, HFTs can be thought of as the continuous analogue of separators connecting null points and several numerical and analytical models have shown that a local O-point structure in the current layer is associated with the dynamic bifurcation of separators and the formation of similar new flux tubes/flux systems as we have found here \citep{Parnell2010,Wilmot-Smith2011,MacTaggart2014,Pontin2015}. We conclude that it is highly likely that such HFT bifurcations would occur dynamically due to the onset of tearing, although the exact nature of this formation is beyond the scope of this investigation.}}

\subsubsection{Flare ribbon signatures}

{{Turning now to the effect of the plasmoids on the flare ribbons,}} the field lines in Fig. \ref{fig:vap_hft} demonstrate that these flux ropes have four {{relevant}} footpoints; two in the straight main sections and two in the hooked ends. Figure \ref{fig:hft_ends} shows the localised nested loops that form in $Q$ at a foot point within a straight section of the ribbon {{where the connectivity of each field line domain has been highlighted. The loop shapes adjacent to the main ribbon outline the quasi-boundary between the newly formed flux tubes and their surroundings. In particular, the loop in Fig. \ref{fig:hft_ends}(a) shows the quasi-boundary between the flux within the green flux tube shown in Fig. \ref{fig:vap_hft}(a) and the adjacent arcade field. The inner loop in Fig. \ref{fig:hft_ends}(b) shows the quasi-boundary between the magenta flux tube shown in Fig. \ref{fig:vap_hft}(b) and the {green flux system which has one half turn less}. Again, we have found analogous nested flux tube formation in models of separator reconnection \citep{Pontin2015}. This closed-loop structure forms directly as a result the plasmoid being located at the junction of four magnetic flux domains (the HFT axis), which facilitates the formation of new flux systems with distinct connectivity. This is fundamentally different from the spiral structure discussed in the following sections that results from twisting up the boundary of two flux domains.}}

Given that the flux tubes that form them map to the center of the flare current sheet, its not clear if only the high $Q$ regions would brighten {{in the flare ribbon}}, or the entire nested region. However, as Fig. \ref{fig:index_comp}(e) to (f) shows the formation of the multiple HFTs and nested flux regions is easily destroyed if the flux rope is moved slightly off the axis of the original HFT. Based on the sensitivity of their positioning, i.e. that they need to form perfectly along a \emph{line in space} (the centre of the HFT), tearing modes of this type seems an unlikely candidate for flare ribbon structure in general. A much more plausible candidate are flux ropes forming on the \emph{surfaces} of high $Q$ defining the borders of the four flux regions. These correspond to oblique tearing modes.

 \begin{figure}
   \centering
   \includegraphics[width=\columnwidth]{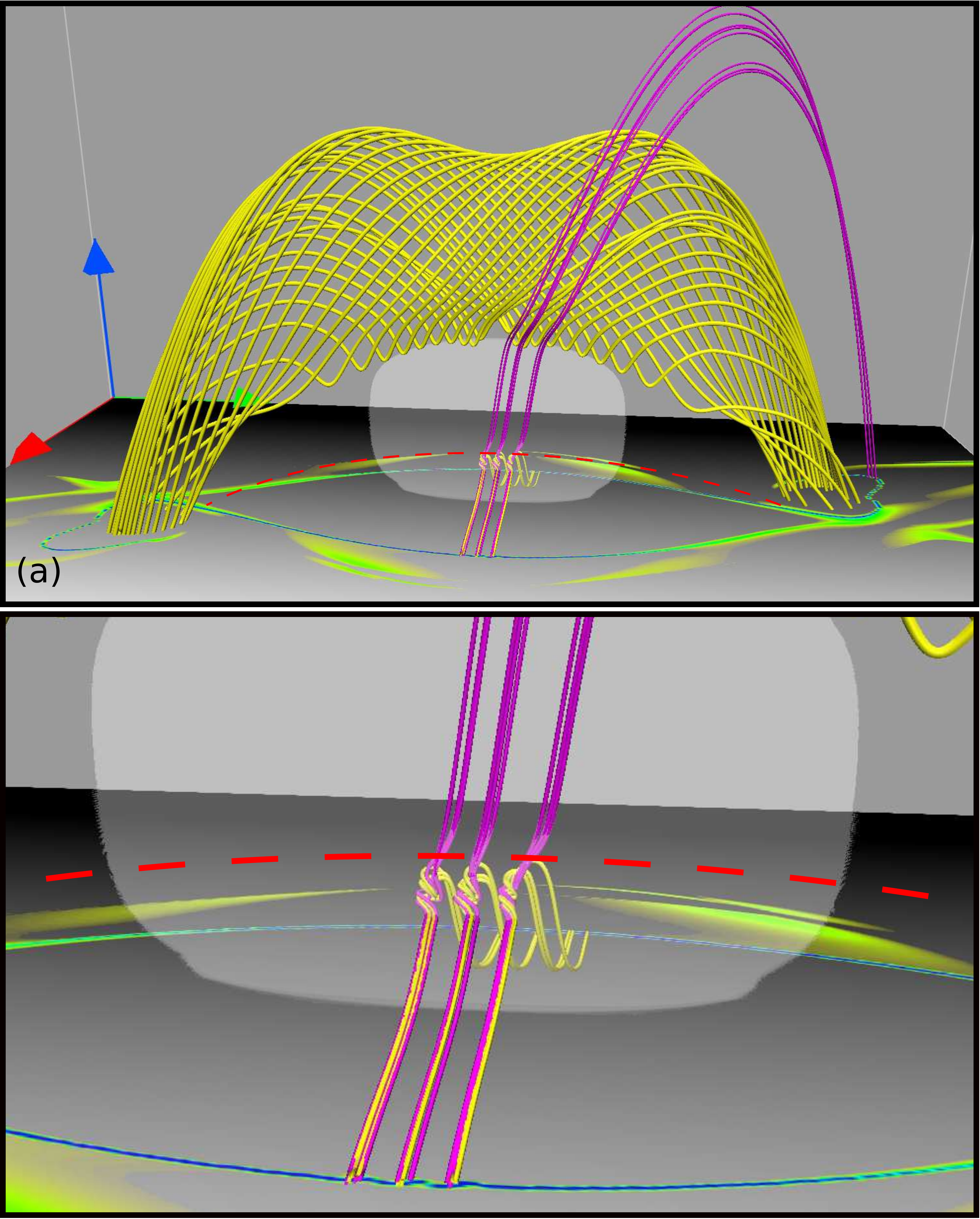}
   \caption{Field lines within three oblique flux ropes formed within the current layer below the HFT. The dashed red line shows the approximate position of the HFT.}
              \label{fig:vap_leg}
    \end{figure}   

 \begin{figure}
   \centering
   \includegraphics[width=\columnwidth]{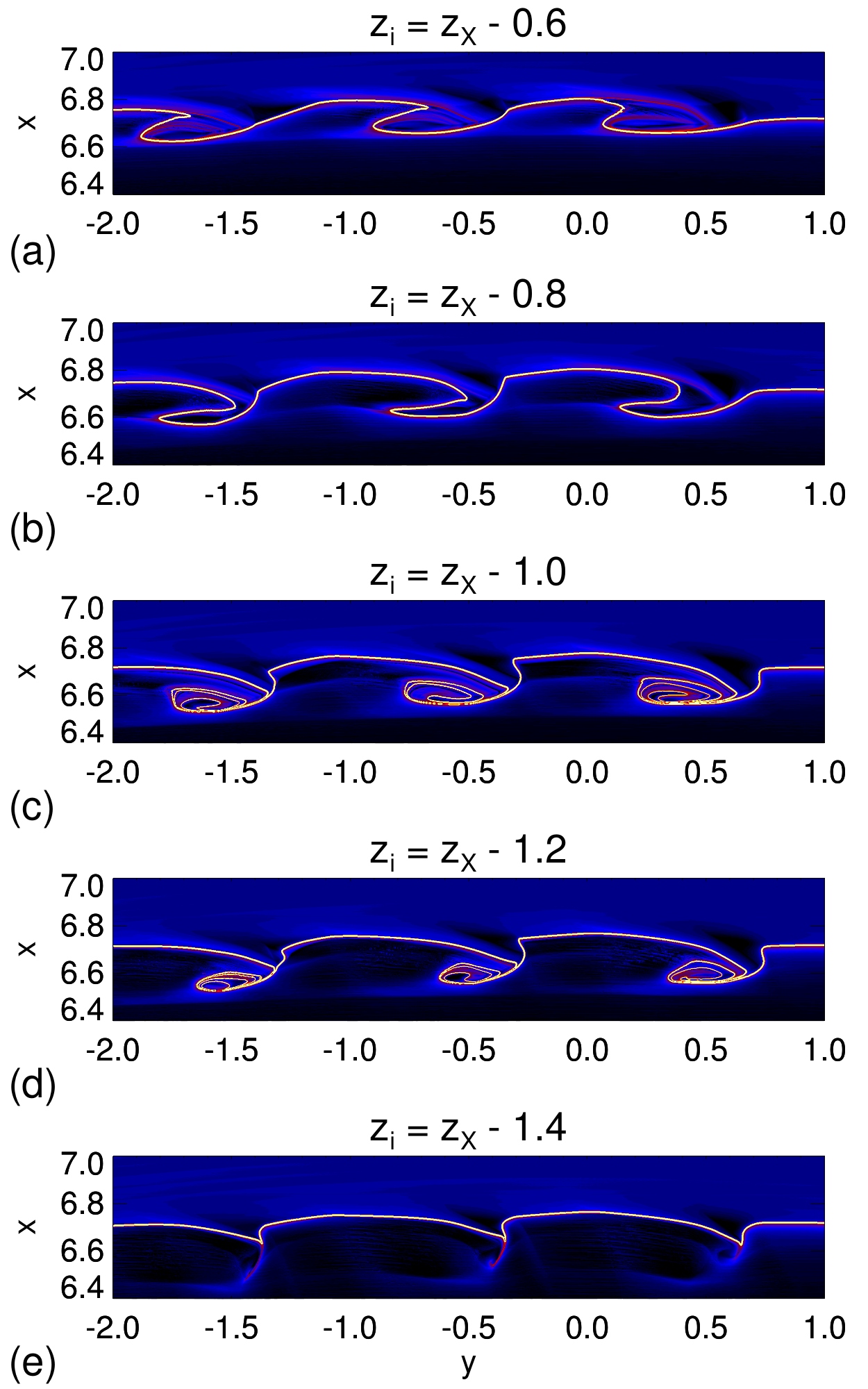}
   \caption{$\log(Q)$ showing the sub-structure formed in the straight section of the flare ribbon by multiple oblique islands ejected downwards from the current layer. Foot points of arcade field lines are above the ribbon, whereas flare loop foot points are below. An animation of panels (a)-(e) is available online showing the evolution. The animation is $1$s long {and begins at $Z_i = Z_x - 0.5$ and ends at $Z_i = Z_x - 1.5$. Between each frame, $Z$ changes by $\Delta Z = 0.1$.}}
              \label{fig:sawtooth}
    \end{figure}

 \begin{figure}
   \centering
   \includegraphics[width=\columnwidth]{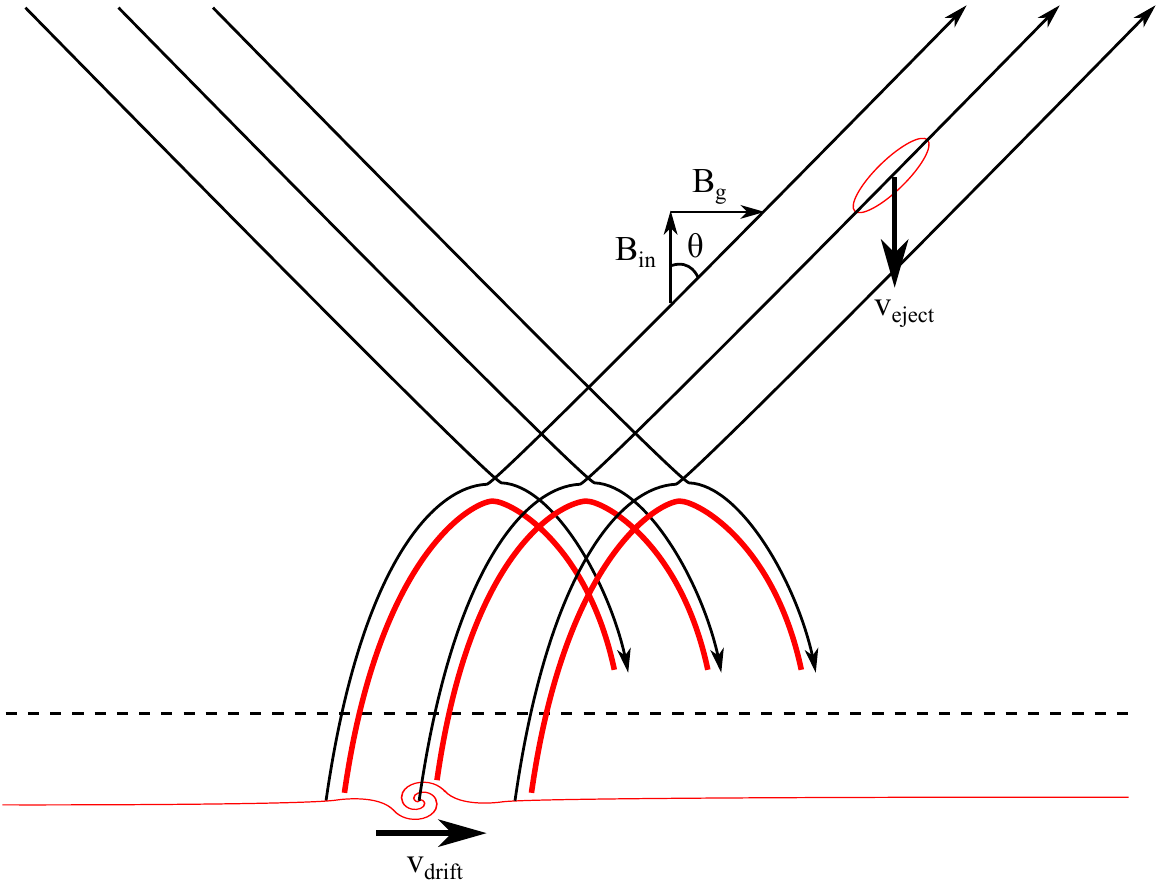}
   \caption{$v_{\text{drift}}$ cartoon showing how as the oblique flux rope is ejected downwards, it slips on to lower field lines leading to a drift of the spiral along the flare ribbon.}
              \label{fig:vdrift}
    \end{figure}
    
 \begin{figure}
   \centering
   \includegraphics[width=\columnwidth]{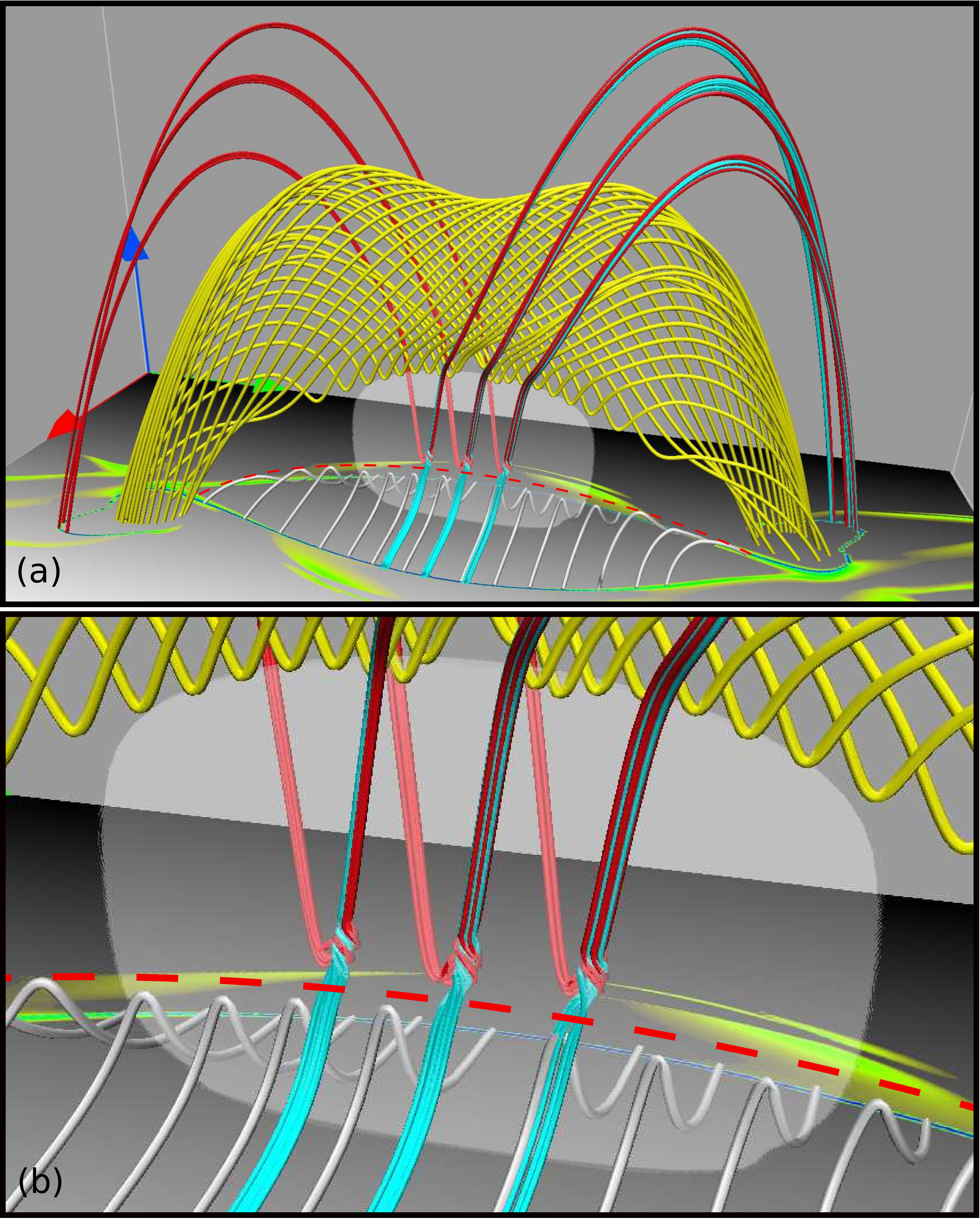}
   \caption{Field lines within three oblique flux ropes formed within the current layer above the HFT. The dashed red line shows the approximate position of the HFT.}
              \label{fig:vap_top}
    \end{figure}

 \begin{figure}
   \centering
   \includegraphics[width=\columnwidth]{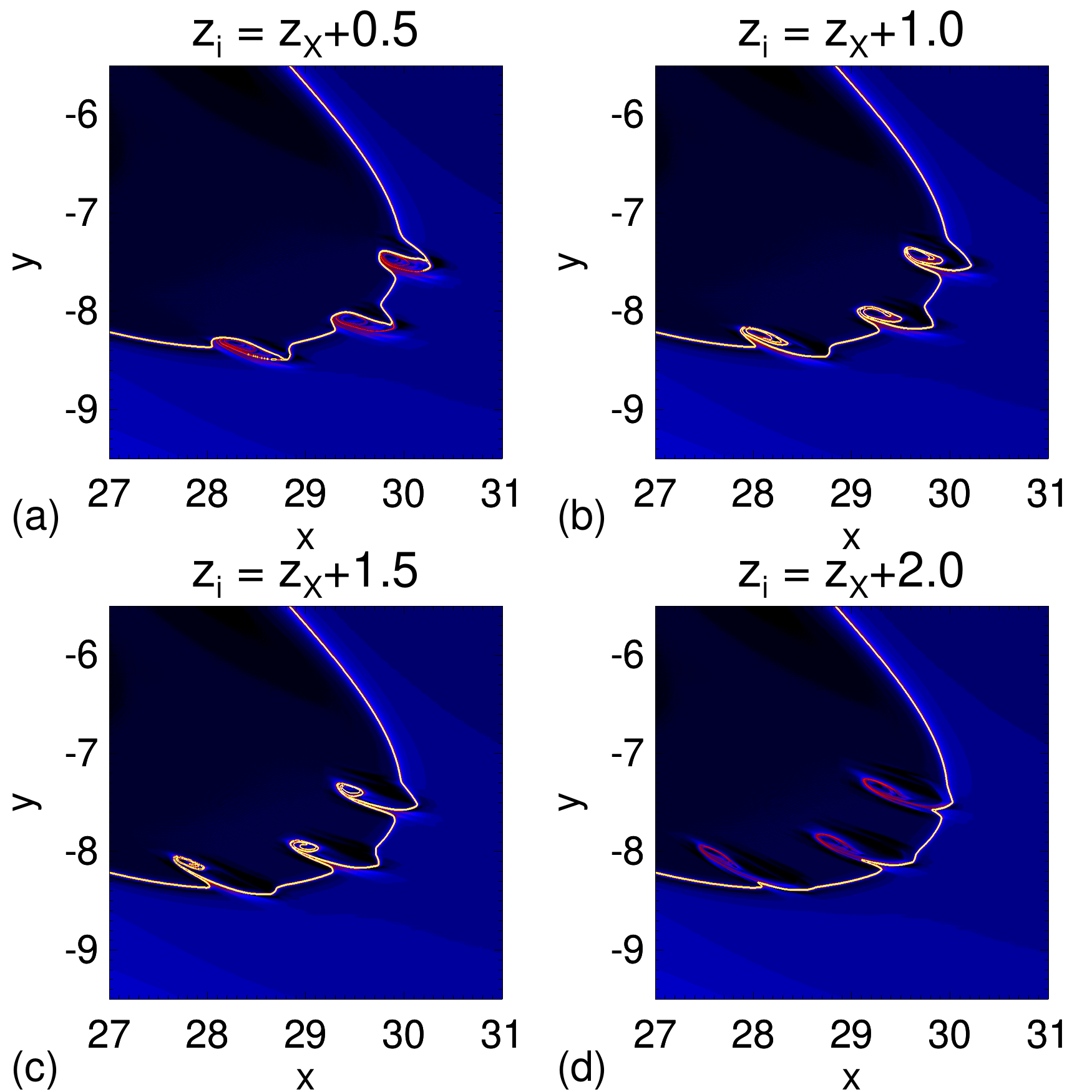}
   \caption{$\log(Q)$ showing the sub-structure formed in the hooked section of the flare ribbon by multiple oblique islands that are ejected upwards from the current layer. Foot points of field lines from the sheath around the main erupting flux rope are towards the top left.}
              \label{fig:3hook}
    \end{figure}

 \begin{figure*}
   \centering
   \includegraphics[width=2\columnwidth]{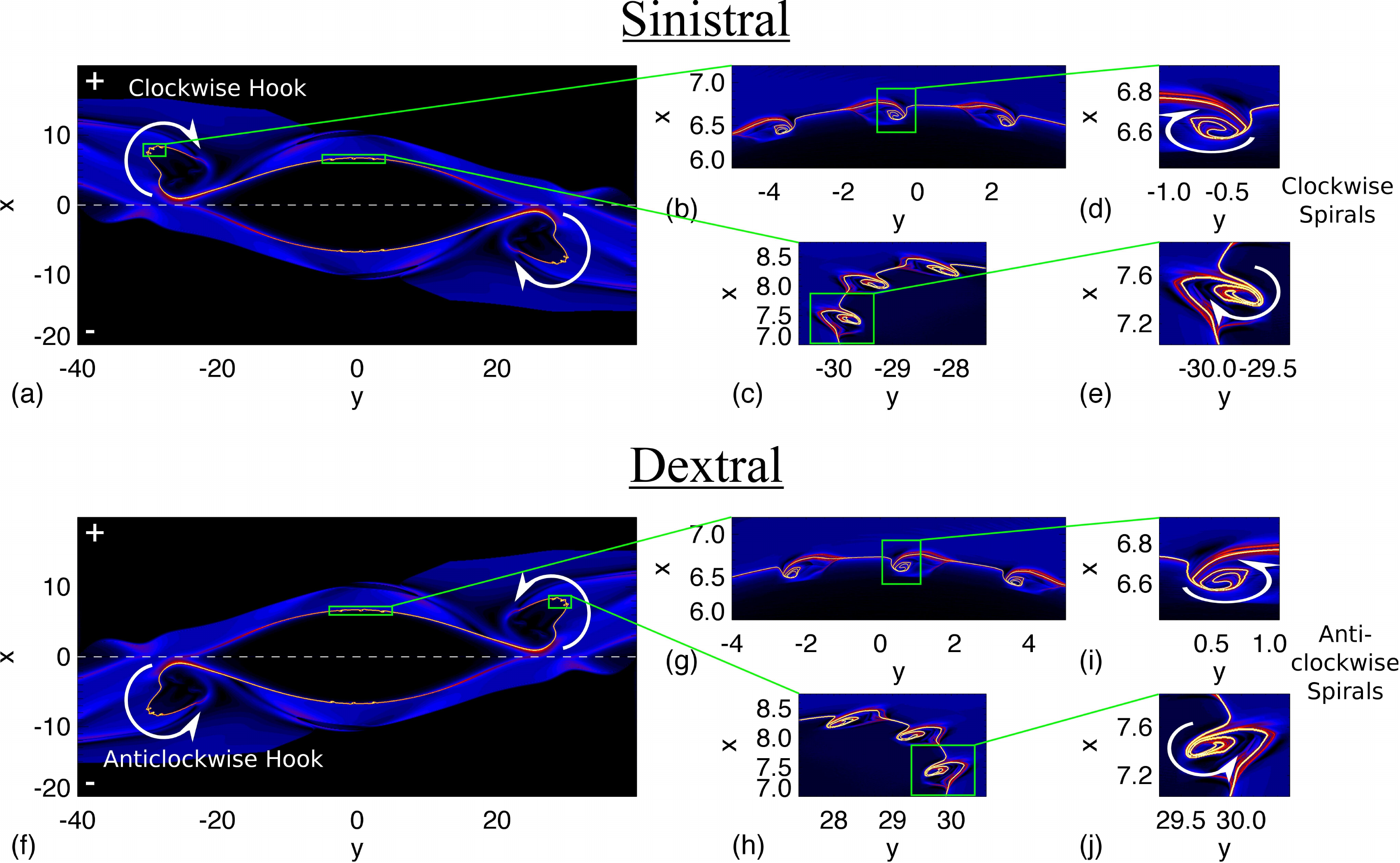}
   \caption{$\log(Q)$ for sinistral {{(a) and dextral (f) flux ropes. (b) to (e) show close up views of the spirals in the boxes shown in (a). Similarly, (g) to (j) show close up views of the spirals in (f). Note in some of the close up views the aspect ratio has been stretched to show the shape more clearly.}} Arrows show the handedness of the hooks/spirals {{while the dashed line shows the PIL}}.}
              \label{fig:comp}
    \end{figure*}

\subsection{Oblique flux ropes}
\label{sec:ob}
\subsubsection{Oblique flux ropes on the arcade boundary}
In a classic zero-guide field current layer, the only resonant surface where tearing can occur is the surface where the field reverses, the scenario considered above. However, once a guide field is introduced the current layer becomes a rotational discontinuity where any flux surface within the current layer can support tearing \citep[e.g.][]{Daughton2011,Huang2016,Edmondson2017}. Flux ropes then form aligned to the local field direction within the resonant flux surface, at some angle to the guide field direction. 

In localised studies of tearing these flux surfaces have no special significance. However, in the context of the global field structure associated with an erupting flux rope, the flux surfaces associated with the two sheet-like QSLs that cross at the HFT are directly associated with the flare ribbons. And since oblique modes form aligned to the \emph{local} field direction, if the flux ropes form on QSL flux surfaces they will twist them up; forming \emph{spirals} in the ribbons themselves. 

Consider the scenario where small flux ropes form on a flux surface within the current layer that is not exactly in its center (i.e., at $x\neq 0$). Then the local field direction with which the flux ropes are aligned when they form is tilted with respect to the guide field (since $B_z\neq 0$ for $x\neq 0$). Let us first consider the case where these flux ropes form with $x>0$ and below the HFT, straddling the QSL flux surface that divides the flare loops from the overlying arcade, i.e. the lower right ``leg'' of the high-$Q$ X shown in Fig. \ref{fig:index}(b). Figure \ref{fig:vap_leg} shows three such flux ropes. Here we choose $x_I = 0.05$, $z_I = z_X(y=0) - 1$, with $y_I = \pm 1,0$ and $c_2 = -2$. At this position the magnetic field is locally aligned at an angle of $\approx 73^{\circ}$ to the $y$-direction, so we set $\theta = 73^{\circ}$ to align the axis of flux rope twist to the local field direction. We note that this angle is somewhat high compared to theoretically expected values \citep[e.g.][]{Daughton2011}. We chose this for illustrative purposes, and would expect similar features for smaller angles corresponding to flux ropes forming nearer the center of the layer.

As expected, the field lines that thread through the small flux ropes form two bundles: the flare loops (yellow) and arcade field lines (magenta), which wrap around one another where they meet below the HFT. Thus, the high-$Q$ layer at the boundary between the two is twisted into a spiral as it maps from the HFT, through the flux ropes and down to the surface. This spiral structure appears in the straight section of the ribbon and is shown in Fig. \ref{fig:sawtooth}(c) for the above choice of parameters. With less twisted flux ropes (i.e., smaller $c_2$), rather than spirals we could recover ``breaking wave''-like ribbon features similar to those shown in the top right of Fig. \ref{fig:iris}. 

Since these small flux ropes form beneath the HFT, they would be expected to be ejected downwards as part of the downflowing plasma from the layer towards the flare loops. We can approximately model this by varying the height of the flux ropes within the current layer -- from very close to the HFT to a distance further below it -- as shown in Fig. \ref{fig:sawtooth}(a) to (e). This reveals that the flare ribbon appears to twist up and then untwist as the flux ropes first encounter the QSL and then move beyond it. Figure \ref{fig:sawtooth} also shows that the spirals have a slight drift along the ribbon. The reason for this drift in our model is shown by the cartoon in Fig. \ref{fig:vdrift}. As the small flux ropes are ejected downwards from the current layer they sample field lines that shift the spiral position progressively further forward. In reality, the ``background'' flare ribbon and the growth of twist within the flux ropes themselves will also be time-dependent, permitting the scenario where spirals form in the ribbon directly (via a small-scale flux rope forming on the QSL directly) before drifting off it. However, if the formation and ejection of oblique flux ropes are still relatively fast then the above predicts drifting spiral structure in the straight sections of the flare ribbons due to oblique modes forming below the HFT.

\subsubsection{Oblique flux ropes on the erupting flux rope boundary}
Consider now the alternative scenario where oblique small flux ropes form above the pre-existing HFT. Such flux ropes can form straddling the section of the QSL flux surface that divides the sheath around the erupting flux rope from the arcade field, i.e. the upper right ``arm'' of the high $Q$ X-shape shown in Fig. \ref{fig:index}(b). Figure \ref{fig:vap_top} shows field lines within three such small flux ropes within the current layer with axes of twist aligned to the local field direction (here we choose $x_I = 0.05$, $z_I=z_X(y=0)+1$, with $y_I = \pm 3, 0$, $c_2 = -2$ and $\theta = 80^{\circ}$). Field lines within the flux ropes from the sheath region are coloured red and those from the arcade are shown in cyan. Forming above the HFT these field lines wrap around one another above the HFT, and consequently the spirals in the flare ribbons occur in the hooked end section of the ribbon far from the flare reconnection site itself. Figure \ref{fig:3hook} shows the associated spirals in $Q$, which also demonstrate a drift and twisting/untwisting as the flux ropes are increased in height, modeling the upward ejection of the flux ropes along with upflowing plasma into the underside of the large-scale erupting flux rope. Again, if our assumption that the flux ropes form and are ejected relatively quickly from the flare current layer is reasonable, then even in a self consistent and evolving field we would expect similar drifting spiral flare ribbon structure to form in the hooked ends of the flare ribbons due to oblique modes above the HFT. The drift in this case is away from the straight section of ribbon, towards the end of the hook.

\subsubsection{Relation between chirality of spirals and hooks}
Finally, one further key prediction relating spirals in flare ribbons to flare current sheet structure is how the chirality of the erupting flux rope relates to the handedness (sense of rotation) in the spirals. To be consistent with the global field reversal across the flare current layer, small-scale flux ropes formed due to tearing must have the same sign of twist as the large-scale erupting flux rope above them (hence our choice of a negative value for $c_2$). As a result the spirals (or breaking waves) have the same sense of rotation as the hooked ends of the ribbon, the nature of which is determined by the chirality of the large-scale erupting flux rope. This is true for spirals both in the straight sections and the hooks themselves. In Fig. \ref{fig:comp} we show the full $Q$ footprint of a sinistral ($b_0=1.7$) and dextral ($b_0=-1.7$) flux rope with three small flux ropes added above and below the HFT on each side of the current layer. Note, here we have spaced the small flux ropes equally in $y$ in this case (with $y_I = \pm3, 0$ for each) but use the same parameter choices as above otherwise. The figure shows that in the sinistral case the ribbons curve clockwise around the foot point of the erupting flux rope with clockwise oriented spirals, whereas in the dextral case the opposite is true, i.e. {\it{the spirals/waves match the hooks}}. Returning to the observation shown in Fig. \ref{fig:iris}, we can see that both the spirals/waves and the hooks do indeed have the same handedness; anti-clockwise in this case.

\section{Discussion}
{{
\subsection{Realism of the model}
In this work we have explored the relation between flare-ribbon fine structure and tearing using a parameterised analytical model. It is important to consider the extent to which the model is representative of a snapshot of a true, dynamic evolution. Despite not being formed by a dynamic evolution, the topology of our background state is exactly that of an eruptive two ribbon flare, i.e. a large-scale flux rope above a current sheet formed at an HFT \citep{aulanier2012}. Based on theoretical studies of reconnection in current sheets with and without a large guide field \citep[e.g.][]{Daughton2011,Huang2016,Edmondson2017} we introduced small-scale flux ropes both in the center of the current layer -- as only occurs in 2D current sheets with zero guide field  -- and also on flux surfaces still within the current layer, but not directly in its center so as to approximate oblique tearing modes. The former led to multiple bifurcations of the HFT and the formation of new (quasi-) flux systems, while the latter twisted up the boundary between the different field line regions creating spiral/breaking wave-like structure in the Q ribbons. 

We emphasise that the analysis presented here explores the possible topological states that {\it{could}} occur as a result of such tearing in the flare current layer. The states which are dynamically realizable will depend on the self consistent evolution both of the large-scale erupting flux rope field and the nature of the instability within the flare current sheet. The multiple bifurcations of the HFT for instance may well only occur transiently as in previous MHD simulations \citep{Wilmot-Smith2007}. However, we believe the spiral structure associated with oblique modes is more robust. Relying only on the formation of small-scale flux ropes at an angle to the guide field within the flare current sheet, which are a ubiquitous feature of the non-linear phase of essentially all simulations of 3D guide field reconnection \citep[e.g.][]{Daughton2011,Wyper2014a,Wyper2014b,Huang2016,Edmondson2017,Stanier2019}. 

To get an idea of how the evolution of the small-scale flux ropes might by related to ribbon dynamics we also moved the flux ropes vertically to simulate their ejection from the current layer. Here we make the assumption that the plasmoids would be ejected rapidly compared to the evolution of the large-scale background field. This is clearly a crude approximation of the actual ejection process, but nonetheless it captures some of the salient features. 2.5D simulations of eruptive flares typically show plasmoids forming and being ejected rapidly compared to the large-scale evolution of the system \citep[e.g.][]{Karpen2012}. In such 2.5D simulations, once the plasmoids grow wider than the current layer they behave ideally as they are advected by the net flow they find themselves in \citep[e.g.][]{Guidoni2016}. However, in 3D the small-scale flux ropes have a finite extent. The main body of the flux rope will become advected by the net flow it finds itself in, however the field lines that map from it then map through the current layer and as a result can continue to change their connections. We have seen similar small-scale flux rope evolution with drifting foot points in our previous simulations of tearing in 3D null point current sheets \citep{Wyper2014b,Wyper2016}. In fact, in 3D theoretically field line connections can change within the flux rope itself as well \citep[e.g.][]{Hornig2003}. Therefore, although crude, we believe that the changes in connections associated with the plasmoid ejection via a simple vertical displacement as we have used should not be too different from what one would find in a true dynamic evolution.}}

\subsection{Interpretation of observations}
The analysis of {{section \ref{sec:ob}}} suggests that oblique 3D tearing modes can contribute to fine structure in flare ribbons. As 3D tearing is known to naturally produce a turbulent spectrum of flux ropes, our analysis also suggests a similar turbulent spectrum of flare ribbon fine structure should exist with the largest most coherent spirals forming the least frequently, perhaps explaining why large spirals are relatively rarely observed. Indeed very large spirals may be associated with the rare ``monster plasmoids" predicted by non-linear tearing theory \citep{Uzdensky10}. Furthermore, the magnetic flux within the spirals/wave-like regions is related to the flux within the small-scale flux ropes. Hence, ribbons traversing regions of intense surface magnetic field strength are likely to have smaller, less observable spiral/wave-like fine structure. The important role played by the guide field in forming oblique tearing modes \citep[e.g.][]{Daughton2011,Edmondson2017,Leake2020} also suggests that flare ribbon structure will vary throughout the life of a flare, as the guide field at the main HFT reduces throughout the eruption.

How then does our model compare with other scenarios put forward to explain flare ribbon fine structure and evolution? \citet{Dudik2016,janvier2013} have made the case that the fast motion of bright points/kernels moving along the flare ribbons are signatures of foot point slipping due to reconnection at the HFT. {{Such slippage is a natural result of the three dimensional flare reconnection process and would occur both in a laminar or a fragmented flare current sheet. However, the preferential brightening of particular foot points/kernels over others implies there must be some inhomogeneity in the reconnection process, either within the reconnection region itself or associated with the coronal loops that are reconnected. As we have shown, spiral or wave-like sub-structure within flare ribbons is a likely observational signature of plasmoids/small flux ropes forming within the flare current sheet. When both clear spiral sub-structure and fast moving bright points/kernels are observed together, this suggests that fragmentation of the current layer is likely the modulating factor in producing the distinct bright kernels associated with the foot point slippage process. That is to say, the two elements of ribbon fine structure are likely intrinsically linked to the fact that the flare current sheet is fragmented. However, particularly in regions of high surface magnetic field strength as discussed above, it is not clear with current observations that all kernels are distinct observational features and do not also include under-resolved spirals. Future high resolution, high time cadence observations from for instance DKIST may be able to test whether some or all bright kernels are simply unstructured loop foot points or contain further spiral structure.}}

\citet{Parker2017} also attempted to explain the wave-like evolution of some flare ribbons \citep{Brannon2015} using a 2D model of tearing with an added velocity shear to create a drift. For the tearing in their model to map to the surface, they envisaged tearing occurring within a current sheet formed along the legs of a flare loop rather than in the flare current layer itself. This is similar to what our model suggests for oblique modes formed on the arcade/flare loop boundary, but with the key difference that the flare current layer in their model is separate to the layer within which they modeled tearing to occur. Here we have shown that when the full three-dimensional structure of the erupting flux rope and flare current sheet are accounted for, tearing modes within the flare current layer itself are able to reproduce the wave-like ribbon structure and potentially also its drift. However, we note that a shortcoming of our model is that our current layer is planar, whereas simulations reveal it to stretch along the QSLs to the flare ribbons \citep[e.g.][]{janvier2013}, so in reality both scenarios may occur.  Indeed, in a similar manner for fine structure occurring in the hooks, small flux ropes need only form somewhere on the QSL surface wrapping over the flux rope. This could be in the flare current layer as we have modeled, or in secondary current layers formed dynamically around the erupting flux rope \citep{Aulanier2019}. The basic premise however is the same.

\section{Conclusions}
The analysis described above leads us to conclude that at least some flare-ribbon fine structure is likely to be related to tearing within the flare current layer. It reveals a direct link between fine structure in the QSLs that align with the flare ribbons, and flare current layer tearing. Our analysis suggests that the dominant contribution to this fine structure comes from oblique tearing modes {{and that where these modes form in the flare current layer is directly related to where they appear in the flare ribbon.}} Plasmoids/flux ropes formed in the reconnection upflow region and that straddle the erupting flux rope/arcade boundary create spiral or breaking wave-like structure in the hooked ends of the ribbons. By contrast, plasmoids/flux ropes formed in the downflow region and that straddle the arcade/flare loop boundary form similar structures in the parallel straight sections of the ribbons. Furthermore, our model predicts that the handedness of the spirals/waves matches the direction of the hooks in the main ribbon itself.

On the assumption that the timescale for plasmoid ejection from the flare current layer is short compared to the timescale of the large-scale flux rope eruption, we also varied the position of the plasmoid flux ropes to crudely model their ejection from the current layer. The spiral evolution produced is remarkably similar to the squirming and breaking wave-like evolution seen in some flare ribbons \citep[e.g.][]{Brannon2015,Dudik2016}, with the spirals/waves exhibiting a drift away from the surface footpoints of the HFT field line; towards the end of the hook in the hooked section, and away from the hook in the straight section. However, we note that the rate of drift would change if we were to relax our simple assumption of a purely vertical ejection. 

The next step is clearly to test these ideas and the predictions of the model against full MHD simulations and observations, work that is currently underway. Such studies may provide a framework for deducing certain properties of the reconnection process on the basis of ribbon fine-structure and its evolution. In this exciting time of high resolution flare ribbon observations from for example IRIS, the New Vacuum Solar Telescope (NVST) and now also the Daniel K. Inouye Solar Telescope (DKIST) it is hoped that this model will spur on further investigation of transient wave-like and spiral flare ribbon structures.

\acknowledgments
We would like to thank Joel Dahlin and Peter Young for stimulating discussions regarding flare reconnection and ribbon signatures. We also thank Peter Young for providing the IRIS movie used to make Fig. \ref{fig:iris}. {{We would also like to thank the anonymous referee for their insightful comments which helped improve our manuscript.}} 
DP acknowledges financial support from STFC through grants ST/N000714 and ST/S000267. 

\appendix
\section{Generalised flux rope field}
\label{ap:A}
The original field from \citet{Titov2009} is given by 
\begin{align}
\mathbf{B}_{fr} = \boldsymbol{\nabla}\times (A_0  \hat{\mathbf{y}}) + b_0 \hat{\mathbf{y}},
\end{align}
with
\begin{align}
A_0 = \frac{x^2}{2} + z+\frac{\epsilon(t)z}{(1+y^2/L_{y}^2)\left(1+z^2/L_z^2\right)},
\end{align}
where $L_y$ and $L_z$ are constants. They chose $b_0=0.2$ and varied $\epsilon(t)$ so that a flux rope formed dynamically. The cusps in this field however are well above the solar surface and the shape of $Q$ on the surface doesn't closely resemble that typically seen in eruptive flares. This field can be generalised such that
\begin{align}
A_0 = \frac{x^2}{2} + z+\frac{\epsilon(t)(z-z_0(y,t))}{(1+y^2/L_{y}^2)\left[1+(z-z_0(y,t))^2/L_z(y,t)^2\right]},
\end{align}
where $L_y$ is constant and the functions $L_z(y,t)$ and $z_0(y,t)$ can be chosen so as to curve the flux rope upwards in its centre and downwards at its ends to place the cusps near the surface and to produce a $Q$ distribution that better resembles the morphology of observed flare ribbons. In this case the $X$ and $O$-point heights are given by
\begin{align}
z_{O,X} = z_{0}(y,t)+L_{z}(y,t)\left\{ \pm\left( \frac{2\epsilon(t)}{1+y^2/L_{y}^2}\left(\frac{\epsilon(t)}{8(1+y^2/L_{y}^2)} -1\right) \right)^{1/2} + \frac{\epsilon(t)}{2(1+y^2/L_{y}^2)}-1 \right\}^{1/2}.
\end{align}
As in \citet{Titov2009}, real values for $z_{O,X}$ occur when $\epsilon(t) \ge 8$ and $|y| < y_{max}$, where
\begin{align}
y_{max}(t) = L_{y}\left(\frac{\epsilon(t)}{8}-1\right)^{1/2}.
\end{align}
$\epsilon(t) = 8$ is therefore the threshold value beyond which a flux rope and HFT form. When $\epsilon(t)>8$ the cusps are then situated at
\begin{equation}
(x_{c},y_{c},z_{c}) = (0,\pm y_{max}(t),z_{0}(y_{max}(t),t)+\sqrt{3}L_{z}(y_{max}(t),t)).
\end{equation}
Here we consider a static flux rope for simplicity, so that $\epsilon$ is constant and we choose
\begin{align}
L_{z}(y,t) &= \beta (1-(y/y_{max})^2)+\gamma, \\
z_{0}(y,t) &= z_{min}-\sqrt{3}\gamma,
\end{align}
where $z_{min}$, $\gamma$ and $\beta$ are constant. $z_0$ is a constant displacement which sets the positions of the cusp points to be
\begin{align}
(x_{c},y_{c},z_{c}) &= (0,\pm y_{max},z_{min}),
\end{align}
while $L_z$ curves the flux rope upwards in its center, above the cusps. We then choose $\epsilon = 16$, which sets $y_{max}=L_y$ to give the form shown in equation \ref{eqn:vect}.

\section{Current sheet field}
\label{ap:B}

 \begin{figure}
   \centering
   \includegraphics[width=\columnwidth]{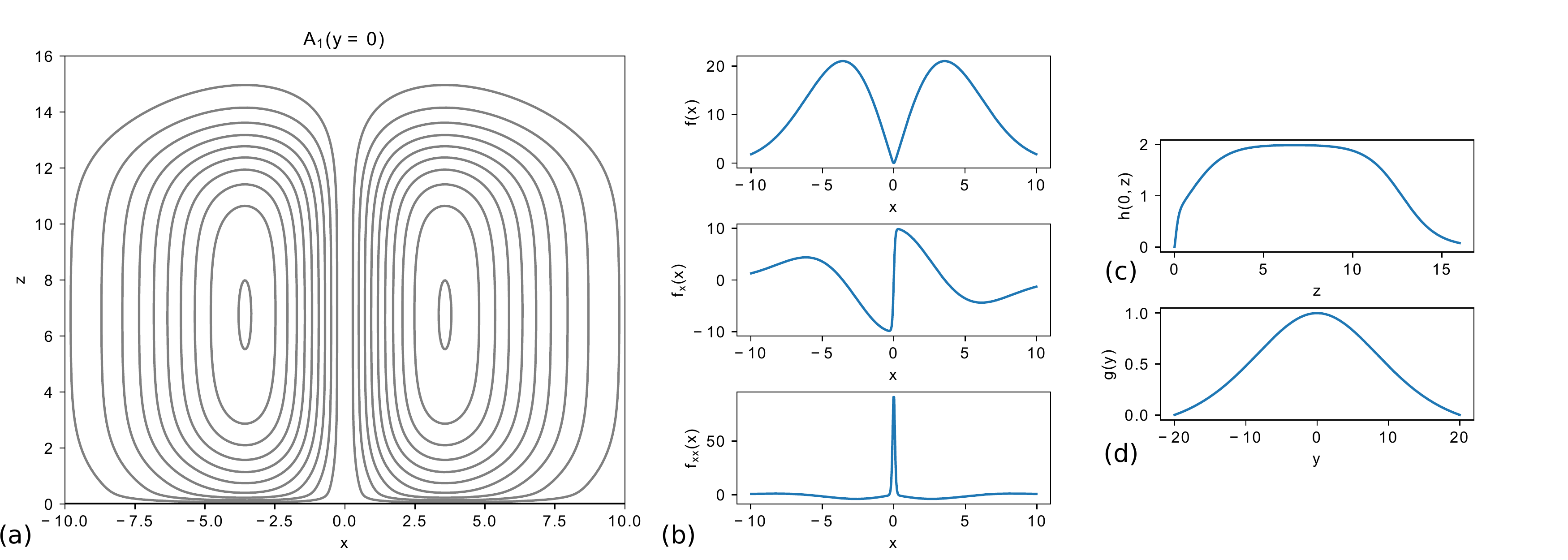}
   \caption{(a) contours of the vector potential for the current sheet field in the $y=0$ plane. (b) $f(x)$ and its derivatives. (c) $h(y,z)$ evaluated at $y=0$. (d) $g(y)$.}
              \label{fig:app1}
    \end{figure}
 
{{$\mathbf{B}_{cs}$ is derived by taking the curl of 
\[
\mathbf{A}_1 =  c_{1} f(x) g(y) h(y,z) \hat{\mathbf{y}},
\]
where $f(x)$, $g(y)$ and $h(y,z)$ are given in equation (\ref{eqn:A1}). Contours of the vector potential ($|\mathbf{A}_1|$) are shown in Fig. \ref{fig:app1}(a), evaluated in the $y=0$ plane. Broadly, this field takes the form of two large rotations of like sign with the HFT sandwiched between them. $f(x)$ controls the $x$ variation of the rotations, with the width of the strong gradient region between the rotations (and therefore the width of the current layer) controlled with $l_x$, and the large-scale extent of the rotations controlled via $k_x$. The line plots in Fig. \ref{fig:app1}(b) show the sharp but continuous gradient in the derivative of $f$ ($f_x$), which sets the field reversal across the current sheet alongside the sharply localised peak in the double derivative of $f$ ($f_{xx}$), which sets the current within the current sheet. $h(y,z)$ is composed of two displaced $\tanh$ profiles centered on $z=z_X(y)$ but offset by $z_c$. $z_c$ therefore defines the vertical length ($\approx 2z_c$) of the current sheet, while $l_z$ sets the rate of drop off of current at the sheet ends through the steepening/flattening of the $\tanh$ profiles. The factor of $\tanh(z/k_z)$ was introduced to guarantee no perturbation of $B_z$ on the surface.  The combined profile of $h(y,z)$ at $y=0$ is shown in Fig. \ref{fig:app1}(c). Finally, the factor $g(y)$ (shown in Fig. \ref{fig:app1}(d)) modulates the strength of the twists so that it peaks at $y=0$ and decays to zero at the cusp points situated at $y=\pm L_y = \pm 20$.}}


\end{document}